\documentclass[aps,preprint,groupedaddress]{revtex4}
\usepackage[dvips]{graphicx}

\newcommand{\ic}{\'{\i}}

\makeatletter
\def\btt#1{\texttt{\@backslashchar#1}}%
\DeclareRobustCommand\bblash{\btt{\@backslashchar}}%
\makeatother

\begin{document}

%%%%%%%%%%%%%%%%%%%%%%%%%%%%%%%%last change%%%%%%%%%%%%%%%%%%%%%%%%
%%\begin{document}
%\renewcommand{\baselinestretch}{1.8}
%\small\normalsize
%
%\vspace*{1.5cm}
%
%\noindent {\bf\textbf{p\,-TYPE SEMICONDUCTING PROPERTIES IN
%LITHIUM-DOPED MgO SINGLE CRYSTALS}}
%
%\vspace*{0.5cm}
%
%\noindent M. M. Tard{\ic}o, R. Ram{\ic}rez, and R. Gonz{\'a}lez
%
%\noindent {Departamento de F\'{\i}sica, Escuela Polit\'{e}cnica
%Superior, Universidad Carlos III,\\
%Avda. de la Universidad, 30, 28911 Legan\'{e}s, Madrid, Spain}
%
%\vspace*{0.5cm}
%
%\noindent Y. Chen\\
%{Division of Materials Sciences, Office of Basic Energy Sciences, SC 13,\\
%U.S. Department of Energy, Germantown, Maryland 20874-1290}
%

%%%%%%%%%%%%%%%%%%%%%%%%%%%%%%%%%%%%%%%%%%%%%%%%%%%%%%%%%%%%%%%%%%%%%%%%
%%%%%%%%%%%%%%%%%%%%%%%%%%%%%%%%%%%%%%%%%%%%%%%%%%%%%%%%%%%%%%%%%%%%%%%%%
							   %% TITLE %%
\title{\bf p\,-TYPE SEMICONDUCTING PROPERTIES IN \\LITHIUM-DOPED MgO SINGLE CRYSTALS}

%%%%%%%%%%%%%%%%%%%%%%%%%%%%%%%%%%%%%%%%%%%%%%%%%%%%%%%%%%%%%%%%%%%%%%
%							    %% AUTHORS %%

\author{M. M. Tard{\ic}o}
\email{mtardio@fis.uc3m.es}
\author{ R. Ram{\ic}rez}
\email{ramirez@fis.uc3m.es}
\author{R. Gonz{\'a}lez}
\affiliation{\em Departamento de F\'{\i}sica, Escuela Polit\'{e}cnica
Superior, Universidad Carlos III,\\
Avda. de la Universidad, 30, 28911 Legan\'{e}s, Madrid, Spain}
\author{Y. Chen}
\affiliation{\em  Division of Materials Sciences, Office of Basic \\
Energy Sciences, SC 13,\\U.S. Department of Energy, Germantown,
Maryland 20874-1290}

%\date{\May 21, 2001}

%%%%%%%%%%%%%%%%%%%%%%%%%%%%%%%%%%%%%%%%%%%%%%%%%%%%%%%%%%%%%%%%%

%\hspace {6.0cm}  {Abstract}
%
%\vspace*{0.5cm}

\begin{abstract}

The phenomenally large enhancement in conductivity observed when
Li-doped MgO crystals are oxidized at elevated temperatures was
investigated by dc and ac electrical measurements in the temperature
interval 250-673 K. The concentration of [Li]$^{0} $centers (Li$^{ +}
$ ions each with a trapped hole) resulting from oxidation was
monitored by optical absorption measurements.

At low electric fields, dc measurements reveal blocking contacts. At
high fields, the $I-V$ characteristic is similar to that of a diode
connected in series with the bulk resistance of the sample. Low
voltage ac measurements show that the equivalent circuit for the
sample consists of the bulk resistance in series with the junction
capacitance connected in parallel with a capacitance, which represents
the dielectric constant of the sample. Both dc and ac experiments
provide consistent values for the bulk resistance. The electrical
conductivity of oxidized MgO:Li crystals increases linearly with the
concentration of [Li]$^{0}$ centers. The conductivity is thermally
activated with an activation energy of (0.70 $\pm $ 0.01) eV, which is
independent of the [Li]$^{0}$ content. The \textit{standard
semiconducting} mechanism satisfactorily explains these results. Free
holes are the main contribution to band conduction as they are trapped
at or released from the [Li]$^{0}$-acceptor centers.

In as-grown MgO:Li crystals, electrical current increases dramatically
with time due to the formation of [Li]$^{0}$ centers. The activation
energy values between 1.3 and 0.7 eV are likely a combination of the
activation energy for the creation of [Li]$^{0}$ centers and the
activation energy of ionization of these centers. Destruction of
[Li]$^{0}$ centers can be induced in oxidized crystals by application
of an electric field due to Joule heating up to temperatures at which
[Li]$^{0}$ centers are not stable.

\end{abstract}


\maketitle
%%%%%%%%%%%%%%%%%%%%%%%%%%%%%%%%%%%%%%%%%%%%%%%%%%%%%%%%%%%%%%%%%%%%%

\newcommand{\figuno}{
\begin{figure}[bh]
\centering
\includegraphics*[scale=0.5]{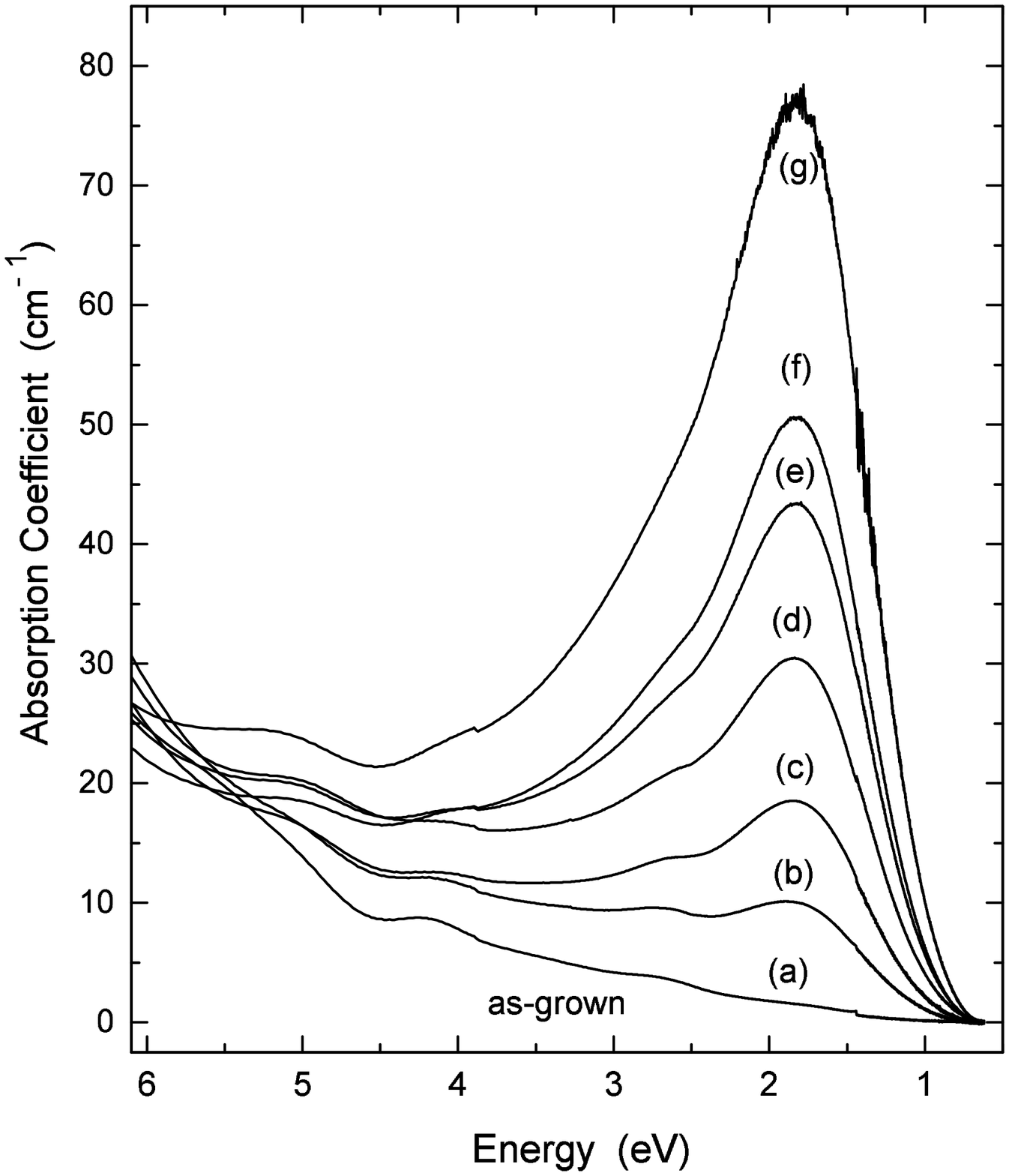}
\caption{Optical absorption spectra of an MgO:Li crystal a) as-grown, and
after oxidation at successively higher temperatures for 30 min at b) 1123 K,
c) 1173, c) 1273 K, d) 1323 K, e) 1373, f) 1273 K, and 1323 K.} \label{fig:one}
\end{figure}
}

\newcommand{\figdos}{
\begin{figure}[tb]
\centering
\includegraphics*[scale=0.5]{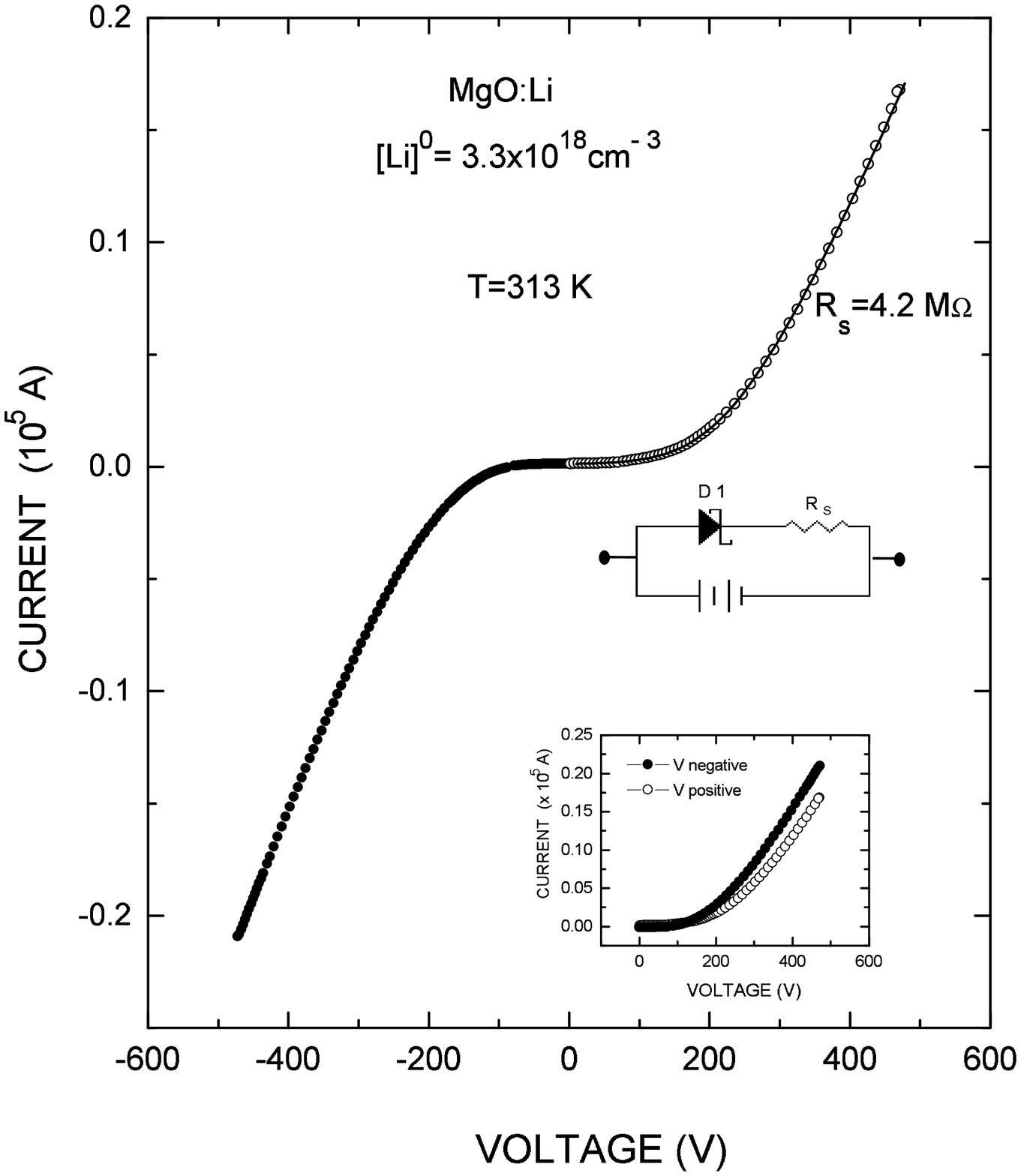}
\caption{Direct current $I-V$ characteristic at 313 K for an MgO:Li crystal
containing [Li]$^{0}$ centers. In the inset the negative and positive parts of
the curve are superimposed.}
\label{fig:two}
%\label{fig:three}
\end{figure}
}

\newcommand{\figtres}{
\begin{figure}[tb]
\centering
\includegraphics*[scale=0.7]{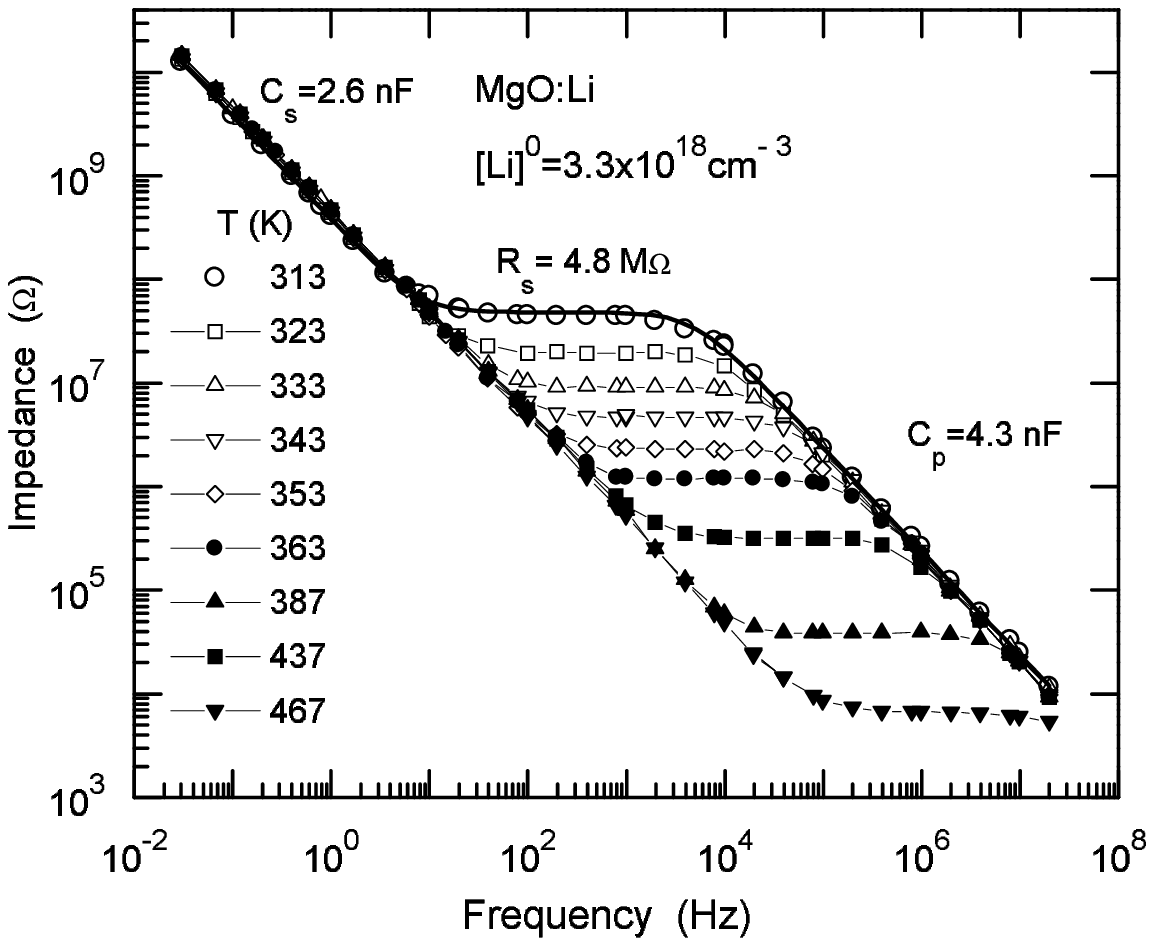}
\caption{Log-log plot of the impedance versus frequency for an MgO:Li crystal
containing [Li]$^{0}$ centers. The solid line represents the best fit of the
experimental points at 313 K to the equivalent circuit.}
\label{fig:four}
\end{figure}
}

\newcommand{\figcuatro}{
\begin{figure}[th]
\centering
\includegraphics*[scale=0.7]{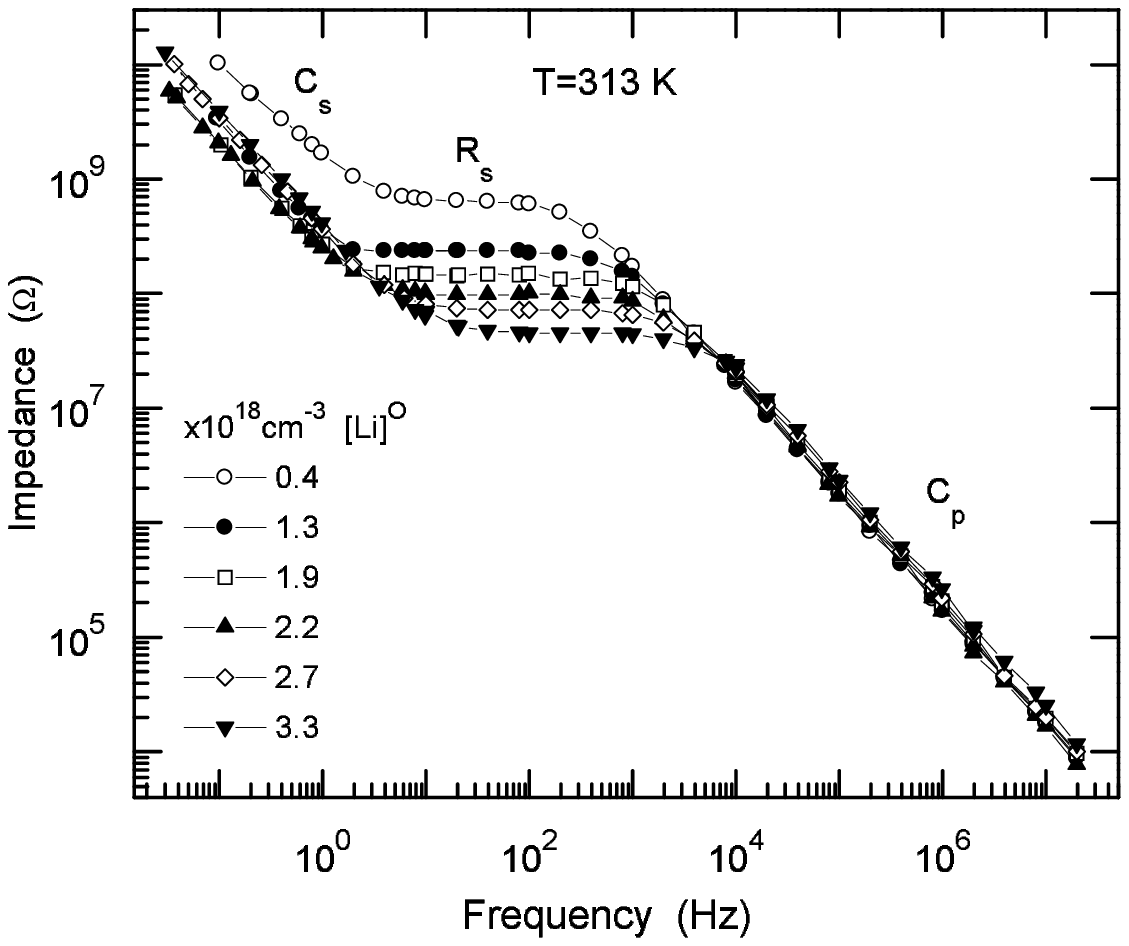}
\caption{Log-log plot of the impedance versus frequency for an MgO:Li crystal
containing different concentrations of [Li]$^{0}$ centers.} \label{fig:five}
\end{figure}
}

\newcommand{\figcinco}{
\begin{figure}[bh]
\centering
\includegraphics*[scale=0.7]{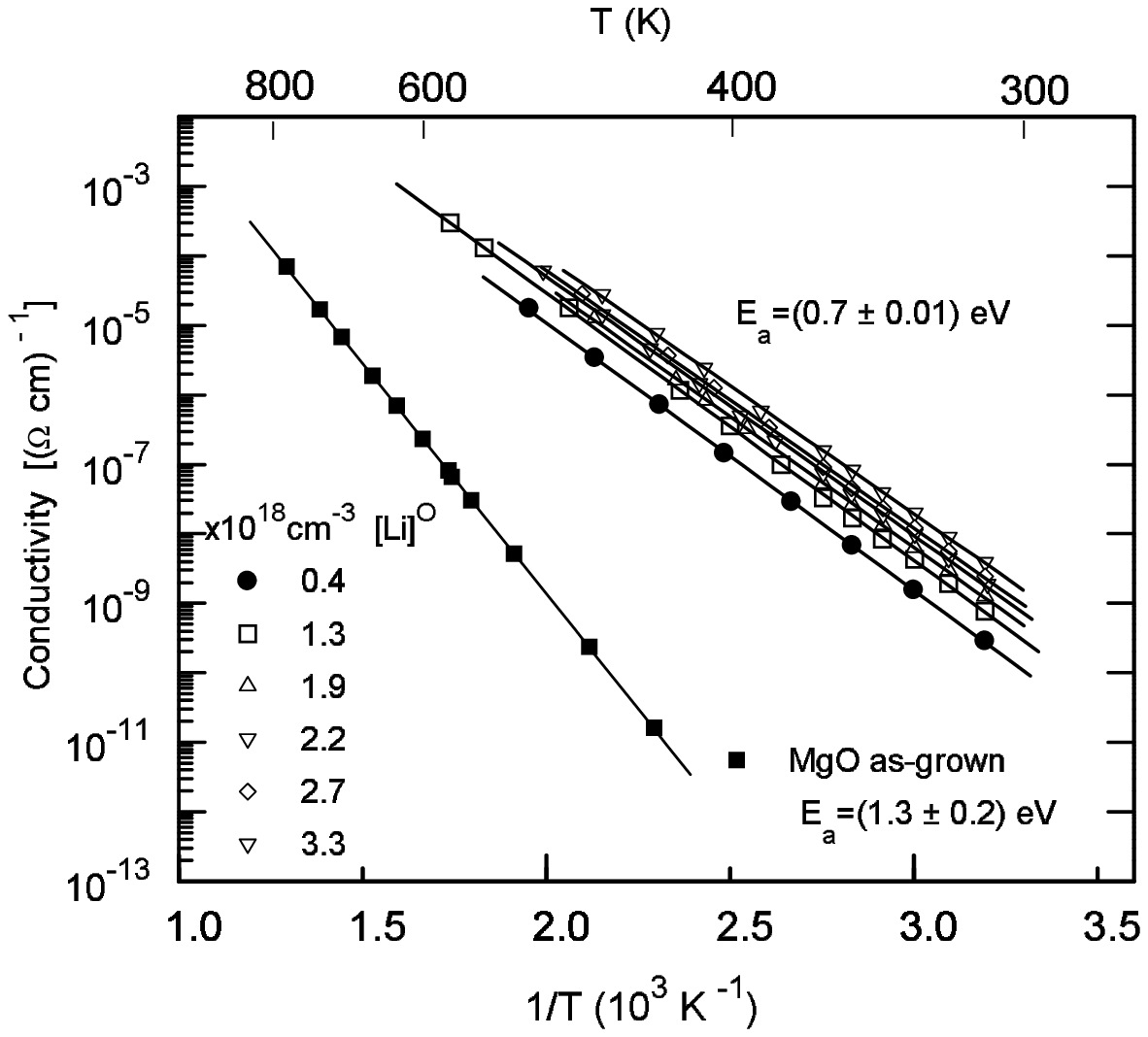}
\caption{Conductivity against $T ^{\mbox{--}1}$ for an as-grown crystal and for
a crystal with different concentrations of [Li]$^{0}$ centers.}
\label{fig:six}
\end{figure}
}

\newcommand{\figseis}{
\begin{figure}[th]
\centering
\includegraphics*[scale=0.5]{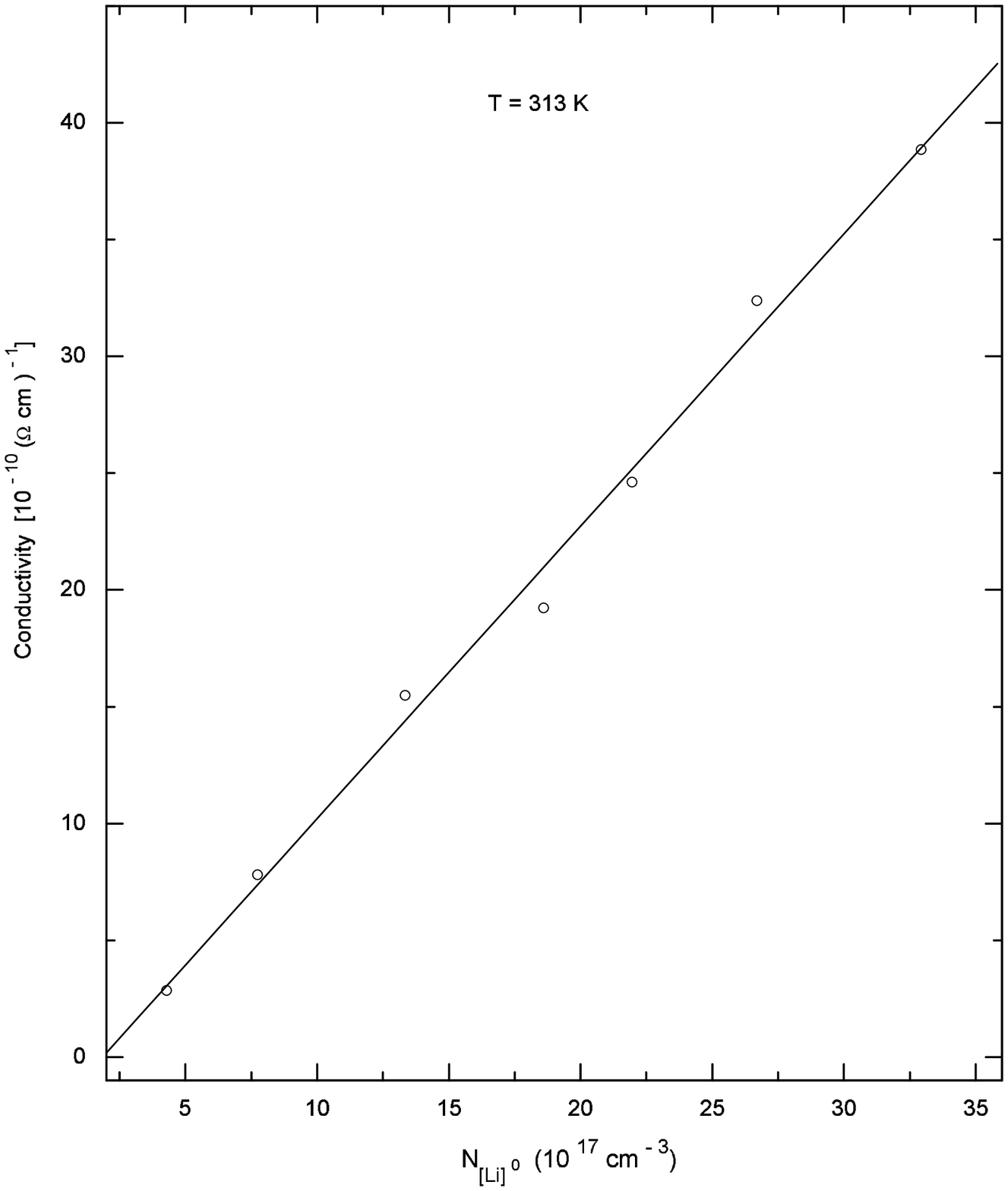}
\caption{Conductivity against concentration of [Li]$^{0}$ centers at 313 K.}
\label{fig:seven}
\end{figure}}

\newcommand{\figsiete}{
\begin{figure}[tb]
\centering
\includegraphics*[scale=0.7]{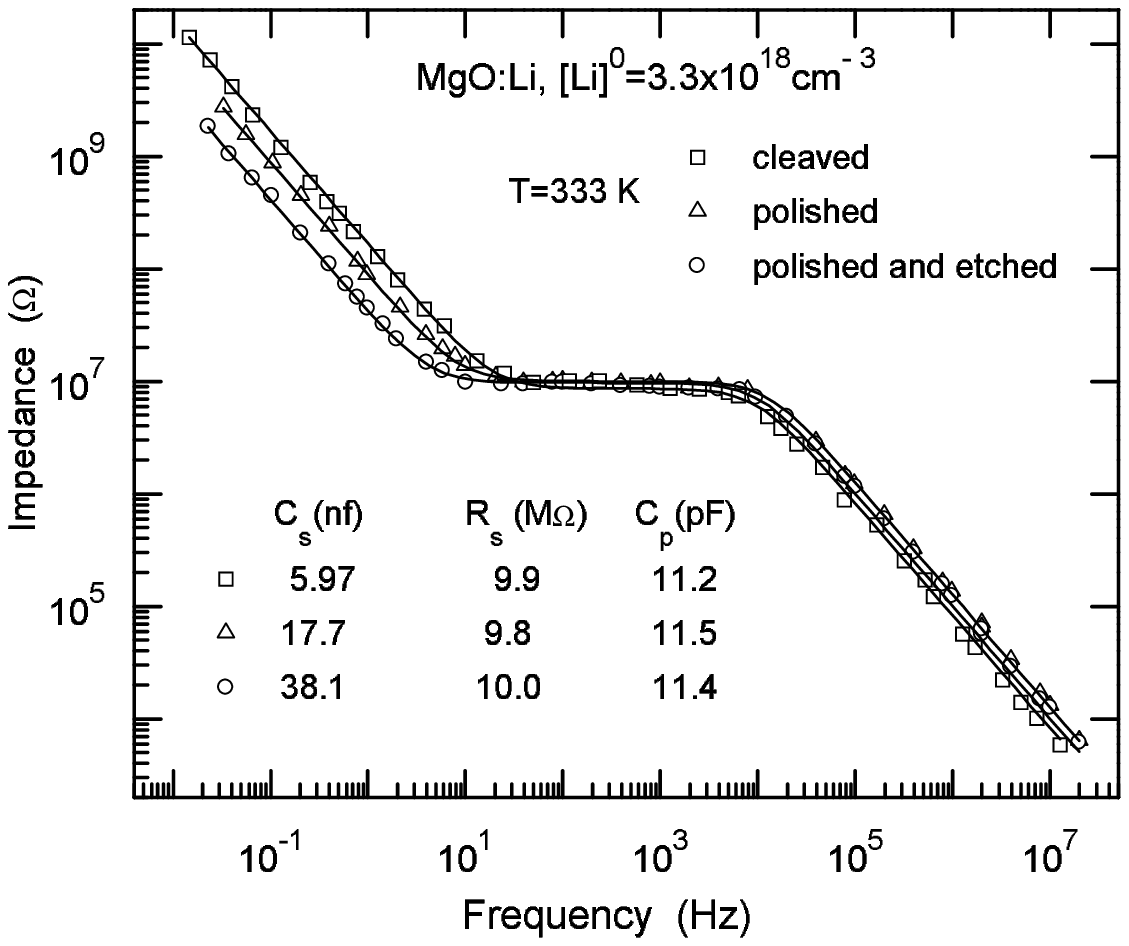}
\caption{Log-log plot of the impedance versus frequency at 333 K for\textbf{
}an MgO:Li crystal containing [Li]$^{0}$ centers and with different surface
conditions.}
\label{fig:eigh}
\end{figure}}

\newcommand{\figocho}{
\begin{figure}[th]
\centering
\includegraphics*[scale=0.5]{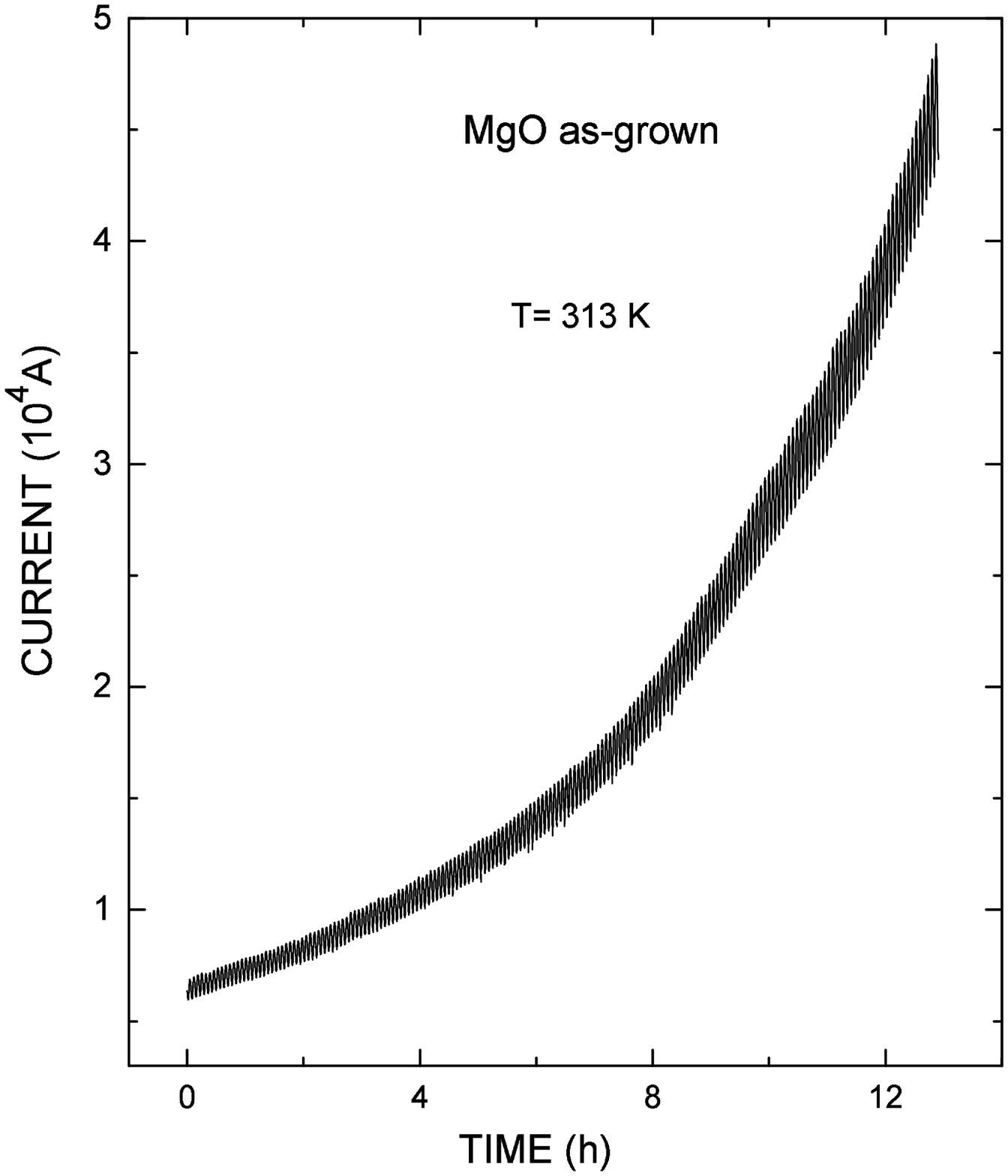}
\caption{Electrical current versus time for an as-grown MgO:Li crystal at
313K.}
\label{fig:nine}
\end{figure}}

\newcommand{\fignueve}{
\begin{figure}[bh]
\centering
\includegraphics*[scale=1.8]{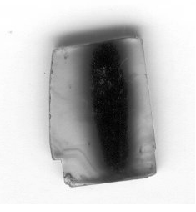}
\caption{Optical photograph of an MgO:Li crystal subjected to an electric
field of 160 V/cm at 1373 K in vacuum for 2 hours. The dark region corresponds
to the area under the electrodes.}
\label{fig:ten}
\end{figure}
}

%%%%%%%%%%%%%%%%%%%%%%%%%%%%%%%%%%%%%%%%%%%%%%%%%%%%%%%%%%%%%%%%%%%%
				%% INTRODUCTION %%
\newpage
\section*{I. INTRODUCTION}

Nominally pure $\alpha $-Al$_{2}$O$_{3}$ and MgO single crystals are excellent
electrical insulators. At room temperature (RT) their electrical
conductivities are 10$^{ - 18}$ and $<$ 10$^{ - 20}$ ($\Omega $cm)$^{ - 1}$,
respectively~\cite{Will:92,Evans:95,Chen:80}. Replacing the host cation with
an aliovalent impurity should in principle result in either electron or hole
conduction, although the presence of other impurities, which act as charge
compensators, might hinder this effect. Even so, a phenomenally large increase
in conductivity at RT or at temperatures of several hundred degrees above RT
was observed in $\alpha $-Al$_{2}$O$_{3}$ and MgO crystals when doped with
magnesium and lithium ions,
respectively~\cite{Chen:80,Tardio2:01,Tardio3:01,Rafa:97}. In both systems
hole-trapped centers are formed after oxidation at high temperatures. These
centers are responsible for the electrical conductivity
enhancement~\cite{Chen:80,Tardio2:01,Tardio3:01,Rafa:97}.

In as-grown MgO:Li crystals, most of the lithium impurities are present in
Li$_{2}$O precipitates~\cite{Narayan:78} and only $ \approx $ 10{\%} as
substitutional Li$^{ +} $ ions;~\cite{Chen3:81} these defects are negatively
charged and are referred to as [Li]$^{ -} $ centers. Oxidation at temperatures
in excess of 1100 K, disperses some of the Li$^{ +} $ ions from the
precipitates resulting in the formation of the paramagnetic [Li]$^{0}$ center
(a substitutional Li$^{ +} $ ion with a hole trapped in one of the six adjacent
oxygen ions)~\cite{Schirmer:76,Abraham1:74,Abraham:76}. This defect absorbs
light at 1.8 eV~\cite{Abraham:76}. Similarly, [Mg]$^{0} $centers
(substitutional Mg$^{ +} $ ion each attended by a hole) are$^{} $present in
oxidized $\alpha $-Al$_{2}$O$_{3}$:Mg crystals, and absorb light at 2.56
eV~\cite{Mohapatra:77,Kroger:83}. (The nomenclature used here follows that
proposed by Henderson and Wertz~\cite{Henderson:68} and subsequently expanded
to cover other defects by Sonder and Sibley~\cite{Sonder:72}. The superscript
refers to the net charge of the defect). The hole is essentially confined to
one lattice site, and together with the distortion it induces in the lattice
is termed a small polaron~\cite{Schirmer:76}.

When a dc voltage was applied at $ \approx $ 373 K to a Li-doped MgO crystal
containing [Li]$^{0}$ centers, semiconducting characteristics, such as
negative differential resistance, self- excited current oscillations and
avalanche breakdown, were observed~\cite{Rafa:97}. An impact ionization
mechanism is responsible for these properties~\cite{Rafa:97}. In the present
study, a thorough characterization of the electrical properties of MgO:Li
crystals was performed in the temperature interval 250-673 K. Both ac and dc
electrical measurements were made. The conductivity dependence on [Li]$^{0}$
concentration and temperature was investigated. The \textit{standard
semiconducting} mechanism satisfactorily explains these results, in contrast
with the Al$_{2}$O$_{3}$:Mg system, where the \textit{small-polaron-motion}
mechanism has been proposed to explain its conductivity. These studies suggest
that other wide-band-gap materials, such as Al-doped SiO$_{2}$, might also
exhibit semiconducting properties at temperatures of several hundred degrees
above RT.

%%%%%%%%%%%%%%%%%%%%%%%%%%%%%%%%%%%%%%%%%%%%%%%%%%%%%%%%%%%%%%%%%%
\section*{II. EXPERIMENTAL PROCEDURE}

The MgO:Li single crystals used in this study were grown by the arc-fusion
technique~\cite{Abraham2:76} using a mixture of 5{\%} Li$_{2}$CO$_{3} $ and
high-purity MgO power from the Kanto Chemical Company. The concentration of
lithium impurities in the resulting crystals was approximately 0.04 at. {\%}
(400 ppm). Samples with an area of about 1.5 cm$^{2}$ and with a thickness of
about 0.1 cm were obtained by cleaving and were chemically polished in hot
phosphoric acid.

As-grown MgO:Li samples were oxidized in flowing oxygen, at
temperatures between 1223 and 1523 K, with the samples placed in a
platinum basket inside an alumina tube inserted in the horizontal,
axial hole of a CHESA furnace. Optical absorption measurements were
performed with a Perkin-Elmer Lambda 19 Spectrophotometer.

For dc measurements, voltage was applied to the crystals with a dc Sorensen
DCS 150-7 voltage source. I-V characteristics were measured with an
electrometer (Keithley 6512) and a voltmeter (HP 34401A). A standard three
electrical-terminal guard technique was used~\cite{Blu:79}. For ac
measurements, a function generator (Wavetek) was used, where available
frequencies range from 10$^{- 4}$ to 10$^{7}$ Hz. When the resistance of the
sample was of the order of the impedance of entrance of the voltmeter, the
current in the circuit was determined from the voltage drop measured in a
resistor in series with the samples. Both the applied voltage and the voltage
in the resistor were recorded with a voltmeter (HP 34401A).

Electrodes were made by sputtering metals with different work
functions (Al, Mg and Pt) onto two opposing lateral surfaces, usually
the largest faces. Sputtering is frequently used to make contacts in
electronic devices because it results in metal films with good
mechanical adhesion and presumably yields ideally clean surface, which
has approximately the same surface-state density as the freshly
cleaved surface. The electrical response was independent of contact
electrode materials, so mostly Al electrodes were used. The
temperature of the sample was monitored with a Chromel-Alumel
thermocouple in direct contact with the sample.

%%%%%%%%%%%%%%%%%%%%%%%%%%%%%%%%%%%%%%%%%%%%%%%%%%%%%%%%%%%%%%%%%%%
\section*{III. EXPERIMENTAL RESULTS}
\subsection*{A. Characterization by optical absorption}

To investigate the electrical conductivity of Lithium-doped MgO crystals, dc
and ac electrical measurements were performed in samples with different
concentrations of [Li]$^{0} $centers in the temperature range 250-673 K. The
[Li]$^{0}$ centers were produced by oxidation at high temperatures. In
addition, the electrical conductivity of as-grown MgO:Li crystals was studied.

\figuno

\subsection*{1. Oxidized MgO:Li crystals}

\subsubsection*{A. Characterization by optical absorption spectroscopy}

In MgO:Li crystals, oxidation at temperatures in excess of 1100 K produces a
broad optical absorption band centered at about 1.8 eV(690 nm) due to
[Li]$^{0}$ centers~\cite{Abraham:76}. This method is very efficient in that it
takes only a few minutes to attain saturation level. Optical-absorption curves
following oxidation for 30 min at increasing temperatures in the same sample
are displayed in Fig. 1. The concentration of [Li]$^{0} $centers can be
determined~\cite{Chen3:81} from the optical-absorption coefficient, $\alpha $,
using Smakula's formula:

%\figuno

\begin{center}
$N$ = 6 $\times $ 10$^{15} \quad f^{ - 1}W\alpha $ (1)
\end{center}

\noindent where the oscillator strength $f$ = 0.1 and the half-width
$W$ = 1.44 eV

\subsubsection*{B.1 Direct current electrical properties}

At low electric fields, dc electrical measurements reveal blocking contacts.
However, at high fields, the reverse bias characteristic is that of a "soft"
leaky barrier. The high field increases the electric field in the metal-MgO:Li
interface, thus, the probability for an electron to tunnel from the metal into
the insulator increases. In this situation, the field emission is more
efficient than the thermionic-field emission. A typical forward-bias
current-voltage ($I-V)$ characteristic is shown in Fig. 2 for a sample with a
[Li]$^{0}$ concentration of $ \approx $ 3.3 x 10$^{18}$ cm$^{ - 3}$. This
characteristic is similar to that of a diode with a series resistance,
$R_{s}$, which corresponds to a blocking contact at one side of the sample and
an ohmic contact at the other side; the series resistance is the bulk
resistance of the sample. The experimental points are plotted as open circles,
and the solid line represents the best fit of the data to the equation:

\begin{equation}
 V = {\frac{{nkT}}{{q}}}\times \ln ({\frac{{I}}{{I_{s}}} }
+ 1) + I\times R_{s}
\end{equation}

\noindent which corresponds to a forward biased diode in series with a
resistance. Here$ n$ is the ideality factor of the junction, $q$ is
the carrier charge, and $I_{s}$ is the saturation current.

For bias voltages in excess of \textit{3kT/q}, the current density should be
proportional to exp(\textit{qV/kT}). This ideal behavior is never observed in
practice, instead the current usually varies as exp(\textit{qV/nkT}). We
obtained our best fitting for values of n $>>$ 1. These high values are related
to both the existence of an interfacial layer and to the recombination of
electrons and holes in the depletion region, which is often important for the
high barriers present in electrical contacts with wide-band-gap insulators (7
eV). Also, the initial part of the characteristic is probably influenced by
the breakdown peculiarities in the reversed direction and, consequently, the
values derived for $n$ and $I_{s}$ are not fully reliable.

Three parameters can be obtained from a fit of the experimental $I-V$
characteristics: the bulk resistance $R_{s}$, the ideality factor $n $of the
junction, and the saturation current $I_{s}$. The values of the last two
parameters are affected mainly by the shape of the low voltage part of the
$I-V$ curve. The value of the series resistance ($R_{s}$ = 4.2 M$\Omega )$
provides an order of magnitude for the sample, as the ac results following
this section will show. The association of the ohmic part of the $I-V$ curve
with the sample resistance is based on the fact that the same value for the
conductivity was determined using samples with different thicknesses and
cross-sections.

As the temperature is raised, the bulk resistance diminishes; it is not
feasible to apply the high voltages needed to break the blocking
contact because of Joule dissipation in the sample. In addition, the
Joule heating at currents higher than 10 mA will significantly change
the concentration of [Li]$^{0}$ centers formed by oxidation.

\figdos

Evidence on the non-homogeneity of the contacts is shown in the insert of
Fig.2. The $I-V$ curve is non-symmetric with respect to voltage polarity,
probably because the size of the electric contact regions that experience
breakdown is different on each electrode. To obtain symmetric curves special
care must be taken to assure that sample preparation and electrode deposition
are identical in both electrodes

As mentioned in the experimental section, we used three metals as
electrodes, Mg, Al and Pt, with widely different work functions: 3.6,
4.2 and 5.6 eV, respectively. The results described in this section are
independent of the types of metal used as electrodes. This indicates
that surface effects dominate the barrier formed at the electrodes.

In conclusion, dc measurements in MgO:Li crystals containing [Li]$^{0}$
centers do not provide information other than whether the material behaves
nonlinearly under the application of a high-voltage stress. In particular, it
is not possible to establish the resistive or capacitive nature of the
material by comparing the magnitude of the electrical current in dc and ac. We
will return to this topic in section \textit{B.2}. Upon application of a
moderate electric field there is a flow of direct current through the sample,
which is ohmic in the high voltage regime, and is governed by the bulk
resistance. The impossibility of controlling the distribution of surface
states present in the insulator as well as electrode effects make the
interpretation of these measurements difficult and sometimes not
quantitatively reproducible. Some of these difficulties can be overcome by ac
electrical measurements.

\subsubsection*{B.2 Alternating current electrical properties}

The experimental set-up and the method of analysis of the experimental values
are similar to those described in reference~\cite{Tardio3:01}. The equivalent
circuit for low voltage ac measurements consists of $R_{s}$ in series with
$C_{s}$ (junction capacitance, which accounts for the blocking nature of the
contacts) and $C_{p}$, which represents the dielectric constant of the sample.
The temperature dependence of these three parameters is illustrated in Fig. 3
where the ac impedance was plotted versus frequency at different temperatures
for a sample with a [Li]$^{0}$ concentration of 3.3 $\times $ 10$^{18}$ cm$^{
- 3}$ (Fig. 3). The sample thickness and the electrode area were 0.10 cm and
0.52 cm$^{2}$, respectively. The solid line is the best fit to the equivalent
circuit at 313 K. The best fit is obtained for\textbf{ }$C_{s}$ = 2.6 nF,
$R_{s}$ = 4.8 $\Omega $, and $C_{p}$ = 4.3 pF, which is consistent with the
dielectric constant of MgO:Li. The basic results are: $C_{s}$ and $C_{p}$ are
practically independent of temperature in the 313-467 K range, whereas the
sample resistance diminishes as temperature increases. The $C_{s}$ value
depends on the quality of the sample surface; we will discuss this later. The
data shown in Figs. 2 and 3 (open circles) were measured in the same sample at
the same temperature; the resulting values for $R_{s}$ in both dc (4.2
M$\Omega )$ and ac (4.8 M$\Omega )$ experiments are in good agreement.

\figtres

We have already mentioned that dc experiments using different sample
geometries confirm that the resistance $R_{s}$ is directly related to the
sample conductivity, thus the dependence of the conductivity on temperature
and [Li]$^{0}$ content can be inferred from $R_{s} $measurements. Fig. 4 shows
the ac characteristics for a MgO:Li sample in which the [Li]$^{0}$
concentration was progressively enhanced by oxidation at increasing
temperatures. After each thermal treatment the concentration of [Li]$^{0}$
centers in the sample was determined and ac measurements were subsequently
made. The resistive part of the ac curve in Fig. 4 depends strongly on the
concentration of [Li]$^{0}$ centers. $C_{s}\approx $ 4.6 nF and $C_{p}\approx
$5.6 pF are practically independent of [Li]$^{0}$ concentration. The exception
is the value of $C_{s}\approx $0.7 nF for the lowest concentration.

\figcuatro

Next, we will address the electrical conductivity and its dependence on
temperature and [Li]$^{0}$ concentration. Fig. 5 shows the Arrhenius plot of
the conductivity for different [Li]$^{0}$ concentrations in the same sample
used in Fig. 4. The slope of the plots is the same indicating that the
conductivity is thermally activated with an activation energy of (0.70 $\pm $
0.01) eV, in good agreement with previous findings~\cite{Chen:80,Rafa:97}. The
parallelism of the straight lines indicates that this value is independent of
[Li]$^{0}$ content. The dependence of the conductivity on [Li]$^{0}$ center
concentration at T = 313 K is shown in Fig. 6. These results clearly show that
there is a linear relationship between conductivity and [Li]$^{0}$ content.

\figcinco

\figseis

In order to investigate the influence of the conditions of the sample surface
on $C_{s}$, $R_{s}$ and $C_{p}$, ac electrical measurements were made in a
sample with a [Li]$^{0} $concentration of 3.3 $\times $ 10$^{18}$ cm$^{ - 3}$.
The two faces of the sample used to sputter the aluminum electrodes were: 1)
cleaved, 2) polished with diamond paste (grain size 5 $\mu $m), and 3)
polished with diamond paste and etched in phosphoric acid. After each step,
aluminum electrodes were sputtered and the sample was electrically measured.
Fig. 7 shows the three log-log plots of impedance versus frequency. While$,
\quad R_{s}$ and$ C_{p}$ remain constant, $C_{s}$, changes significantly. This
result is not surprising, the first two parameters are related to the sample
resistivity and to the dielectric constant of the sample, respectively, which
are not affected by the quality of the sample surface. On the other hand, the
junction capacitance, $C_{s}$, depends on the characteristics of the Al-MgO:Li
interface. These results provide clear evidence that the random distribution
of surface states controls the interface barrier, and, as was already
observed, $C_{s}$, must be independent of the work functions and the electron
affinities of both the metal and the insulator, which are in contact.	 In
addition, the impossibility of making an ohmic contact in one of the
electrodes makes it very difficult to characterize the barrier properties
associated with the [Li]$^{0}$ concentration.

\figsiete

\section*{2. As-grown $\mathrm{MgO:Li}$	 crystals}

Previous experiments at room temperature have found the conductivity of
as-grown Li-doped crystals to be much lower than that of oxidized crystals
containing stable [Li]$^{0} $centers, but much higher than in undoped MgO
crystals~\cite{Rafa2:97}. In the present work, electrical conductivity
measurements were performed in as-grown MgO:Li crystals between 400 and 800 K.

Fig. 8 shows the time evolution of the current in an as-grown MgO:Li sample at
313 K when a low electric field of 100 V/cm is applied. Initially, the current
was of the order of 0.1 $\mu$A and very slowly increases with time. Current
oscillations are superimposed on the sublinear steady increase of the current.
These oscillations are due to oscillations of the sample temperature, with a
period of about 5 min; their amplitude increases as the current increases.
After several hours the current increase becomes more pronounced, and given
enough time, the sample experiences electrical breakdown. The current increase
is attended by the emergence of \textit{blue regions} in the area under the
electrodes. Optical absorption spectra of these regions show an absorption
band which peaks at 1.8, eV similar to that found in oxidized MgO:Li crystals
containing [Li]$^{0}$ centers. This observation is a clear indication that
upon application of a low electric field, [Li]$^{0}$ centers are created,
which are responsible for the current increase. The enhancement of the
oscillation amplitude is a result of a higher amount of carriers due to a
larger concentration of [Li]$^{0}$

\figocho

The temperature dependence of conductivity for an as-grown MgO:Li sample was
determined using ac measurements after an electric field was applied for 20
minutes. The Arrhenius plot is shown on the left side of Fig. 5. The resulting
activation energy is 1.3 eV, which is much larger than the 0.7 eV obtained for
oxidized crystals. However, an electric field of 100 V/cm was applied for
different time intervals at the same sample and consequently the electrical
current significantly increased. After each time interval the activation
energy was again determined. The activation energy was observed to diminish
with increasing time, after several hours it reaches a value of 0.7 eV,
indicating that when the [Li]$^{0}$ concentration induced by the application
of the electric field is sufficiently high, the activation energy in a
non-oxidized crystal approaches that of the oxidized crystal. The activation
energy values between 1.3 and 0.7 eV are likely a combination of the
activation energy for the creation of [Li]$^{0}$ centers and the activation
energy of ionization of these centers.

To get further evidence of the production of [Li]$^{0}$ centers by application
of an electric field, an as-grown MgO:Li crystal was subjected to an electric
field of 160 V/cm at a temperature of 1373 K for 2 min. The experiment was
performed in vacuum to diminish the oxygen partial pressure. Fig. 9 shows the
coloration of the area under the electrode. The optical absorption spectrum of
this region shows an absorption band at 1.8 eV, confirming the presence of
[Li]$^{0}$ centers. In the area outside the electrodes, the crystal exhibits
the amber coloration characteristic of as-grown MgO:Li crystals. The reverse
effect can also be induced by the application of an electric field. In an
oxidized crystal discoloration of localized regions of the area under the
electrodes was observed for currents in excess of 10 mA. At these high
currents, Joule heating will heat up those regions to temperatures at which
the [Li]$^{0}$ centers are not stable.

\fignueve

\section*{IV. DISCUSSION}

The results presented in previous sections show that the electrical
conductivity in oxidized MgO:Li crystals increases linearly with the
concentration of [Li]$^{0} $ centers. The conductivity is thermally
activated, with an activation energy of (0.7 $\pm $ 0.01) eV, which is
independent of the [Li]$^{0} $content.

The thermally activated behavior of conductivity can be explained by any of
three different mechanisms: small polaron
motion~\cite{Emin:69,Frolich:54},impurity
conduction~\cite{Pollak:61,Semic.K:89}, or standard semiconducting
behavior~\cite{Friedman:63}. The predictions of the first mechanism have been
shown to be consistent with the results of dc and ac electrical measurements
in Al$_{2}$O$_{3}$:Mg crystals containing [Mg]$^{0}$ centers. However, in
spite of the similarities between Al$_{2}$O$_{3}$:Mg and lithium-doped MgO
crystals, it is highly unlike that the \textit{small polaron motion mechanism}
is responsible for the conductivity behavior in MgO: Li crystals: although
\textit{bound} polarons are currently used to explain the optical absorption
band at 1.8 eV associated with [Li]$^{0} $centers~\cite{Schirmer:76}, in
contrast with the Al$_{2}$O$_{3}$:Mg system, there are neither experimental
evidence nor theoretical calculations~\cite{Schirmer:76} that justify the
existence of a \textit{free} polaron in MgO with $V$-type defects such as
[Li]$^{0}$ centers. \textit{Impurity conduction} may be ruled out because this
mechanism predicts a strong dependence of the activation energy with the
concentration of [Li]$^{0 }$centers. Fig. 5 shows that the activation energy
is the same over a broad range of [Li]$^{0} $centers. Hence, we are left with
the \textit{standard semiconducting mechanism.}

The temperature dependence of the hole conductivity ($\sigma _{p} = q
\cdot p \cdot \mu _{p} )$ results from the combined effect of the
variations in the hole concentration, $p$, and the hole mobility,
\textit{$\mu $}$_{p}$. The electron charge is denoted by $q$. For band
conduction, the main contribution is due to the change in the
concentration of free holes as they are trapped at or released from
the [Li]$^{0}$-acceptor centers. The variation with temperature of the
hole mobility is relatively small and depends on the scattering
mechanism, which usually varies as a power of the temperature, and it
is independent of the acceptor concentration for concentrations that
are not too large.

The concentration of holes $p$ is given by:

\begin{equation}
\label{eq1}%
p(T) = {\frac{{N_{V} (N_{A} - N_{D} )}}{{2N_{D}}} }\exp (-
\phi / kT)
\end{equation}

\noindent where $N_{A}$ and $N_{D}$ are the concentrations of acceptor
and compensating impurities, respectively, $N_{V}$ is the effective
density of states in the valence band, and\textit{ $\phi $} is the
acceptor ionization energy.

The holes are created by	 ionization of the major neutral acceptors:
[Li]$^{0}$ centers

\begin{equation}
\label{eq2} %
\mathrm{[Li]^{0} \rightleftharpoons [Li]^{-}+h}
\end{equation}

Applying the mass-action law
\begin{equation}
\label{eq3}%
 K_{{\left[ {Li} \right]}^{0}} = {\frac{{N_{{\left[ {Li}
\right]}^{ -}} \cdot p}}{{N_{{\left[ {Li} \right]}^{0}}}} },
\end{equation}

\noindent we obtain:

\begin{equation}
\label{eq4}%
p(T) = 2{\frac{{N_{{\left[ {Li} \right]}^{0}}}}
{{N_{{\left[ {Li} \right]}^{ -}} }} } \cdot N_{V} \exp ( - \Delta
E_{{\left[ {Li} \right]}^{0}} / kT)
\end{equation}

\noindent where:

 $\Delta E_{{\left[ {Li} \right]}^{0}} = (E_{Li_{Mg}^{ -}}	 - E_{V}
)_{thermal} = 0.7eV,$ if there is not an activation barrier;$^{13}$
here E$_{V}$ is the energy at the valence band edges.

The effective density of states in the valence bands is given by$^{22}$

\begin{equation}
\label{eq5}
N_{V} = \left( {{\frac{{2\pi m_{p}^{ *}	 kT}}{{h^{2}}}}}
\right)^{{\raise0.7ex\hbox{${3}$} \!\mathord{\left/ {\vphantom {{3}
{2}}}\right.\kern-\nulldelimiterspace}\!\lower0.7ex\hbox{${2}$}}}
\end{equation}

\noindent where $k$ is Boltzmann constant, $h$ the Plank constant, the
effective hole mass and $T$ the absolute temperature.

Substituting in equation (\ref{eq4}) yields

\begin{equation}
\label{eq6}%
 p(T) = 2{\frac{{N_{{\left[ {Li} \right]}^{0}}}}
{{N_{{\left[ {Li} \right]}^{ -}} }} }\left( {{\frac{{2\pi
m_{p}^{\ast}  k}}{{h^{2}}}}} \right)^{{\raise0.7ex\hbox{${3}$}
\!\mathord{\left/ {\vphantom {{3}
{2}}}\right.\kern-\nulldelimiterspace}\!\lower0.7ex\hbox{${2}$}}}T^{{\raise0.7ex\hbox{${3}$}
\!\mathord{\left/ {\vphantom {{3}
{2}}}\right.\kern-\nulldelimiterspace}\!\lower0.7ex\hbox{${2}$}}}\exp
( - \Delta E_{{\left[ {Li} \right]}^{0}} / kT)
\end{equation}

Finally the holes conductivity $\sigma _{p}$ is given by:

\begin{equation}
\label{eq7}%
\sigma _{p} (T) = 2q{\frac{{N_{{\left[ {Li} \right]}^{0}}}}
{{N_{{\left[ {Li} \right]}^{ -}} }} }\left( {{\frac{{2\pi
m_{p}^{\ast}  k}}{{h^{2}}}}} \right)^{{\raise0.7ex\hbox{${3}$}
\!\mathord{\left/ {\vphantom {{3}
{2}}}\right.\kern-\nulldelimiterspace}\!\lower0.7ex\hbox{${2}$}}}T^{{\raise0.7ex\hbox{${3}$}
\!\mathord{\left/ {\vphantom {{3}
{2}}}\right.\kern-\nulldelimiterspace}\!\lower0.7ex\hbox{${2}$}}}
\cdot \mu (T) \cdot \exp ( - \Delta E_{{\left[ {Li} \right]}^{0}} / kT)
\end{equation}

This equation gives a linear dependence of the conductivity
 with the concentration of [Li]$^{0}$ in accordance with our
experimental results (Fig. 6).

In addition, the (ln\textit{$\sigma $})-1/$T$ curves displayed in Fig.
5 are straight lines for the temperature range (250-673 K)
investigated and for different [Li]$^{0}$ concentrations. Our
electrical measurements were made at temperatures lower than the Debye
temperature of MgO ($\Xi = $743 K). At these temperatures, the
scattering of holes by the lattice should be more important than the
scattering by impurities that dominates at $T \ll \Xi $. Assuming that
the temperature dependence of the hole mobility is controlled by a
\textit{lattice scattering} mechanism, the hole mobility decreases
with increasing temperature as $T ^{ - 3 / 2}$. This results in a
canceling of the $T^{3 / 2}$ dependence of the carrier concentration
in Eq. (\ref{eq6}) and, consequently, in a linear relationship between
ln\textit{$\sigma $} and 1/$T$, in agreement with the results of Fig.
5. It is worth noting, however, that a temperature dependence of the
preexponential term in Eq. (\ref{eq7}) from $T ^{ - 1}$ to $T ^{ - 2}$
would produce a sufficiently small curvature in the (ln\textit{$\sigma
$} vs 1/$T)$ curves as to go unnoticed.

\section*{V. SUMMARY AND CONCLUSIONS}

Optical absorption measurements at 1.8 eV were used to monitor the
concentration of [Li]$^{0} $centers in oxidized lithium-doped MgO
crystals. Direct current and alternating current electrical
measurements were made to investigate the electrical conductivity of
MgO:Li samples containing different concentrations of [Li]$^{0}
$centers in the temperature interval 250-673 K.

At low electric fields, dc measurements reveal blocking contacts. At
high fields, the $I-V $characteristic is similar to that of a diode
(corresponding to a blocking contact at one side of the sample and an
ohmic contact at the other side) connected in series with the bulk
resistance of the sample. Surface states and electrode effects make it
difficult to interpret these measurements.

Low voltage ac measurements reveal that the equivalent circuit for the sample
consists of the bulk resistance, $R_{s}$ in series with $C_{s}$ (symbolizes
the blocking contacts) connected in parallel with a capacitance $C_{p}$, which
represents the dielectric constant of the sample. Both dc and ac experiments
provide consistent values for the bulk resistance. In the temperature interval
313-467 K, $R_{s}$ decreases with increasing temperature, whereas $C_{s}$ and
$C_{p}$, do not change significantly. The electrical conductivity of oxidized
MgO:Li crystals increases linearly with the concentration of [Li]$^{0}$
centers. The conductivity is thermally activated with an activation energy of
(0.70 $\pm $ 0.01) eV, which is independent of the [Li]$^{0}$ content. These
experimental results are in agreement with the predictions of the
\textit{standard semiconducting mechanism}. The main contribution to band
conduction is related to the concentration of free holes as they are trapped
at or released from the [Li]$^{0}$-acceptor centers. In spite of the
similarities between lithium-doped MgO and Al$_{2}$O$_{3}$:Mg crystals, the
mechanism responsible for the conductivity associated with the presence of
hole-trapped centers in both systems is different: \textit{standard
semiconducting} mechanism and\textit{ small polaron motion},
respectively~\cite{Tardio2:01,Tardio3:01}.\textit{Free} polarons in MgO with
$V$-type defects such as [Li]$^{0}$ were neither observed nor theoretically
postulated~\cite{Schirmer:76}.

In as-grown MgO:Li crystals subjected to an electric field of 100
V/cm, the current was observed to increase dramatically with time, and
in the area under the electrodes blue regions appeared. Optical
absorption measurements show that the coloration is associated with
the presence of [Li]$^{0}$ centers. The temperature dependence of the
conductivity shows that the activation energy varies between 1.3 and
0.7 eV depending on the length of time the electric field has been
applied to the sample. After a sufficiently long time the activation
energy approaches that of oxidized crystals containing [Li]$^{0}$
centers. Activation energy values larger than 0.7 eV are likely a
combination of the activation energy for the creation of [Li]$^{0}$
centers and the activation energy of ionization of these centers.

Both creation and destruction of [Li]$^{0}$ centers can be induced by
application of an electric field. In an oxidized crystal, currents in
excess of 10 mA discolor localized regions of the area under the
electrodes due to Joule heating up to temperatures at which the
[Li]$^{0}$ centers are not stable.

\section{ ACKNOWLEDGEMENTS }

Research at the University Carlos III
was supported by the CICYT of Spain. The research of Y.C. is an
outgrowth of past investigations performed at the Solid State Division
of the Oak Ridge National Laboratory.

\end{document}